# Travaux pratiques sur les réseaux locaux de type WIFI utilisant des simulations numériques des phénomènes de propagation des ondes électromagnétiques

Demontoux F.[1], Miane J.L.[2]

[1] IUT GEII Talence et Laboratoire PIOM-ENSCPB, 16. av. Pey Berland 33607 Pessac, f.demontoux@enscpb.fr

[2] Laboratoire PIOM-ENSCPB, 16. av. Pey Berland 33607 Pessac, jl.miane@enscpb.fr

**RESUME** La technologie de réseau sans fil WiFi trouve de plus en plus d'applications dans le domaine industriel que ce soit pour l'échange d'informations entre les personnes ou les équipements. Il a l'avantage de permettre une grande flexibilité du réseau mais nécessite des précautions d'installation pour être efficace (débit, fiabilité…). L'installateur d'un tel réseau doit posséder des connaissances pour installer et configurer les matériels (points d'accès, cartes wifi) mais aussi des connaissances en propagation des ondes électromagnétiques qui véhiculent les informations. L'enseignement de cette technologie nécessite des travaux pratiques qui doivent permettre à l'étudiant de mettre en œuvre ces connaissances. Le travail que nous présentons retrace les différents moyens de calculs que nous avons mis en œuvre dans le cadre de travaux pratiques pour permettre à l'étudiant de visualiser et ainsi de mieux maîtriser les phénomènes de propagation impliqués dans le fonctionnement d'un réseau WiFI.

**Mots clés** : Travaux pratiques, WiFi, simulation numérique, propagation, recouvrement de cellules, transmission

## 1 INTRODUCTION

La technologie de réseau sans fil WiFi trouve de nombreuses applications dans tous les domaines ; qu'ils soient privés (domotique) ou professionnels (réseau local industriel, échange d'informations). L'enseignement de ces réseaux nécessite des travaux pratiques. Ces derniers doivent aborder les notions qui permettront à l'étudiant d'être capable d'installer et de configurer un tel réseau.

Ce TP a été décomposé en deux parties. La première partie traite de l'installation et de la configuration de la carte WiFi et du point d'accès (chiffrement, canal de transmission, préambule long ou court…). Il traite également de l'étude du réseau ad-Hoc, en mode ad-Hoc ou en mode infrastructure (zone de couverture, routage…). L'utilisation d'un logiciel de capture de trame (LinkFerret) permet aussi l'étude de différents types de trames (données, contrôle, gestion…).

La spécificité de ces réseaux est leur support de transmission qui est l'air. Ainsi le deuxième TP doit permettre à l'étudiant d'acquérir des connaissances en propagation des ondes électromagnétiques. Le but du travail que nous présentons a été de concevoir ce TP pour permettre à l'étudiant d'aborder ces notions. Notre démarche a été d'utiliser un logiciel de simulation des phénomènes électromagnétiques. Nous l'utilisons comme un outil de compréhension des contraintes liées à la propagation des ondes (atténuations, réflexions, recouvrements de cellules…) [1] qui conduiront l'étudiant à effectuer certains choix lors de l'installation d'un réseau WiFi (position des points d'accès, sécurité du réseau…).

Le travail que nous présentons s'inscrit dans le cursus d'enseignement de la licence professionnelle réseau SARI (Systèmes automatisés - Réseaux Industriels) de l'IUT GEII (IUTA- Université Bordeaux 1). Son introduction dans le cadre des travaux pratiques télécommunication d'IUP GEII et de maîtrise EEA (Université Bordeaux 1) dirigés par JL Miane est en cours.

## 2 PRESENTATION DU LOGICIEL DE SIMULATION

Le logiciel que nous utilisons est le logiciel HFSS (High Frequency Struture Simulator) de la société ANSOFT. L'utilisation de cet outil dans le cadre de recherches scientifiques au sein du laboratoire PIOM (Physique des interactions Ondes Matières) nous a permis d'obtenir de la société ANSOFT des licences gratuites pour l'enseignement.

Ce logiciel est basé sur la méthode des éléments finis [3]. C'est une méthode de résolution des équations de Maxwell qui nécessite le découpage de l'espace en tétraèdres sur lesquels s'applique l'algorithme de résolution.

Le but de notre enseignement n'est pas d'aborder la résolution numérique des équations de Maxwell. Cependant une présentation succincte de cette méthode est introduite dans le sujet du TP.

Un des avantages de ce logiciel est de posséder une interface utilisateur sur plateforme Windows XP conviviale qui permet une rapide prise en main [2].

Lors du TP les étudiants sont amenés à effectuer des modifications sur des projets existants (propriétés de certains matériaux…) à lancer des calculs et à interpréter les résultats. La nature et la durée des TP (une séance de trois heures pour cette partie) imposent que le temps de calcul de chaque étude ne soit pas trop long et que ces calculs puissent être effectués sur des ordinateurs « classiques » (512 Mo de mémoire RAM). Malheureusement certains problèmes abordés peuvent nécessiter des temps de calcul et des mémoires importants. Nous verrons dans notre étude que des mesures ont dû être prises pour répondre à ces exigences.

## 3 SIMULATION DE LA PROPAGATION DU CHAMP ELECTROMAGNETIQUE EMIS PAR UN POINT D'ACCES WIFI

### 3.1 Présentation du problème simulé

La première étude que les étudiants sont amenés à faire repose sur l'étude du champ émis par un point d'accès WiFi. Le principe de cette simulation est de faire observer les phénomènes de propagation des ondes WiFi (2400 MHz) dans une pièce.

A l'intérieur de cette dernière se trouvent des cloisons d'épaisseur et de nature différentes (plâtre, béton, bois….) ainsi que du mobilier (armoires en acier…).

Le volume d'une pièce moyenne est d'environ 30m$^3$. La longueur d'onde dans l'air à 2400 MHz est de 12.5 cm. Si nous utilisons 10 cellules de calcul par longueur d'onde il nous faudra environ 6 millions de cellules. Les contraintes de temps de calcul et de mémoire nous obligent à nous limiter au maximum à 60 000 cellules. La solution retenue a été de limiter notre problème à un problème pratiquement à 2 dimensions en limitant la hauteur de la pièce. Cette simplification du problème impose aux objets d'être représentés comme ayant la même hauteur que la pièce. Elle simplifie aussi les phénomènes de propagation. Nous verrons toutefois que les simulations obtenues permettent, par exemple, une étude précise des phénomènes de chemins multiples ou de recouvrement de cellules.

La figure 1 représente la géométrie que nous avons simulée. La pièce fait 16m$^2$ (4mx4m). Nous y avons placé des cloisons et des armoires métalliques.

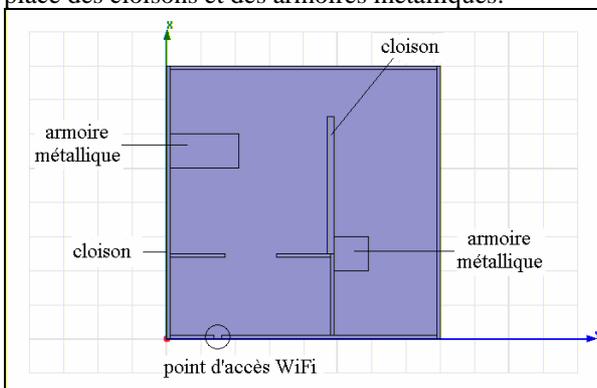

*fig 1 : géométrie représentée*

### 3.2 Champ EM émis par le point d'accès

La première simulation consiste à représenter le champ électrique émis par le point d'accès (Figure 2).

Une observation plus précise de ces résultats (zoom, animation en phase…) permet d'observer la longueur d'onde de l'onde, les phénomènes de chemins multiples, de réflexion (armoire métallique…) ou d'atténuation (cloisons).

Elle permet à l'étudiant d'appréhender les problèmes de « couverture » d'un point d'accès et donc les limitations de la taille d'une « cellule » Wifi. Nous rappelons ainsi que la première mesure de sécurisation d'un réseau wifi est la maîtrise du rayonnement des points d'accès qui peut empêcher une personne extérieure à un bâtiment (par exemple) d'avoir accès au réseau.

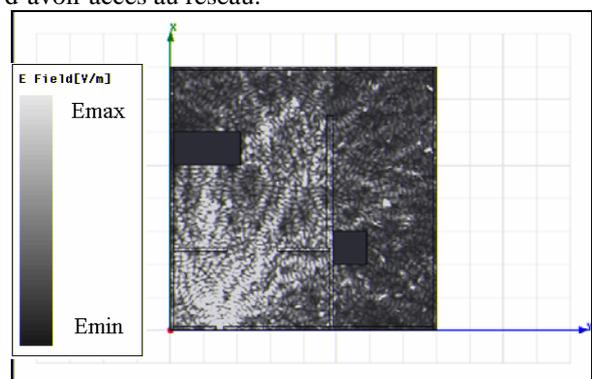

*fig 2 : champ électrique – er=3-0.03j*

### 3.3 Puissance émise par le point d'accès et sensibilité des détecteurs

Un autre paramètre que nous pouvons observer est la puissance émise par le point d'accès. Elle permet de mieux comprendre les problèmes liés à la sensibilité des détecteurs des cartes Wifi utilisées.

Le réglage de l'échelle maximale des couleurs au seuil de détection de la carte permet de visualiser la zone « couverte ». Le changement de seuil de sensibilité (Figure 3 et 4) montre l'importance de ce facteur sur la zone de couverture du réseau.

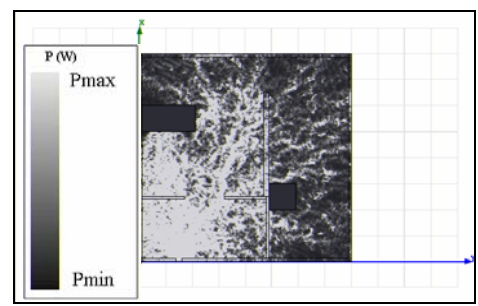

*fig 3 : Puissance observée # 1 – représentations en fonction du seuil de détection*

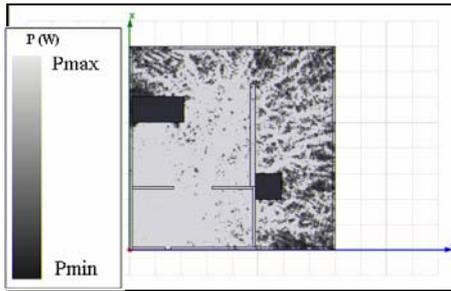

*fig 4 : Puissance observée #2– représentations en fonction du seuil de détection*

L'étudiant aura également la possibilité de comparer ces résultats avec des mesures de champ rayonné par le point d'accès qui lui permettront de tracer une cartographie de la zone « couverte » par le point d'accès (utilisation d'un analyseur de spectre équipé d'un dipôle de réception ou ordinateur portable équipé d'une carte WiFi).

3.4    Transmission entre le point d'accès et une station.

Il est aussi possible de calculer le coefficient de transmission entre le point d'accès et une station du réseau wifi (figure 5) et ainsi définir avec précision la qualité de transmission du signal..

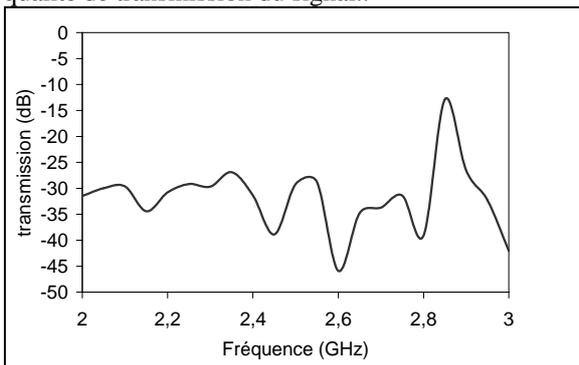

*fig 5 : transmission entre deux points du réseau*

L'étudiant pourra aussi tracer la variation de ce coefficient de transmission en fonction, par exemple, de l'épaisseur d'une cloison ou de sa permittivité (figure 6) ou tracer la cartographie de champ électrique en fonction de la permittivité des cloisons (Figures 7 et 8).

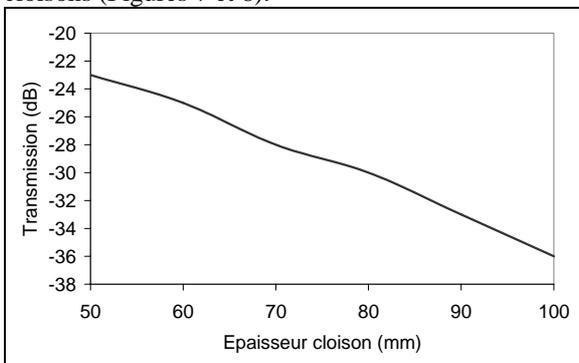

*fig 6 : transmission en fonction de l'épaisseur des cloisons*

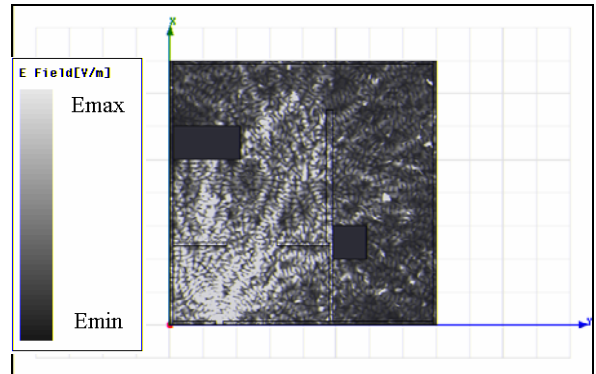

*fig 7 : champ électrique – er=3-0.03j*

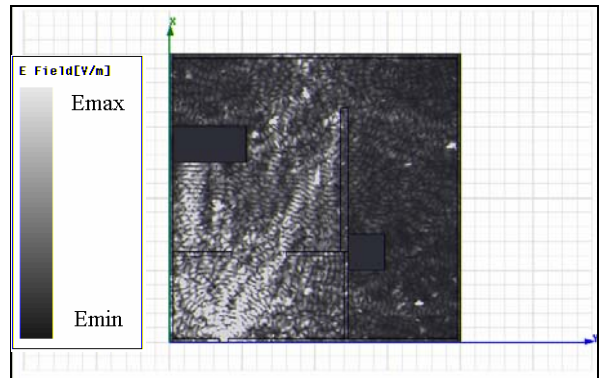

*fig 8 : champ électrique – er=5-0.2j*

3.5    Recouvrement de cellules

Le protocole Wifi 802.11f permet l'interopérabilité entre les points d'accès au moyen du protocole de gestion des handovers IAPP (Inter – Access Point Protocol) et donc la continuité de service. Son utilisation n'est pas encore généralisée et l'utilisateur ne peut pas passer d'une cellule à une autre en maintenant sa connexion.

Toutefois les réseaux wifi doivent souvent se baser sur une architecture de réseau ambiant qui repose sur l'utilisation de cellules voisines qui se recouvrent partiellement pour assurer, par exemple, la couverture complète de tout un bâtiment. Les résultats des figures 9 à 11 présentent ce phénomène en simulant l'utilisation de deux points d'accès.

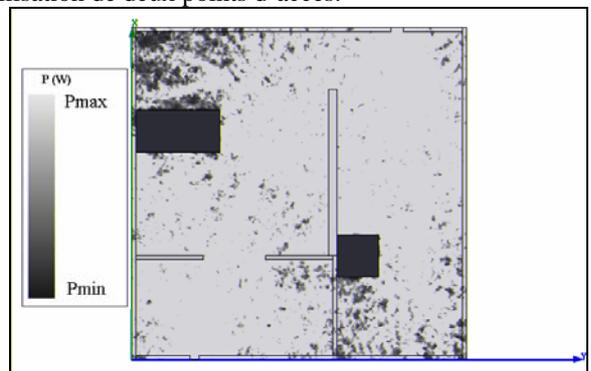

*fig 9 : Recouvrement de cellule et hand-over utilisation des deux sources*

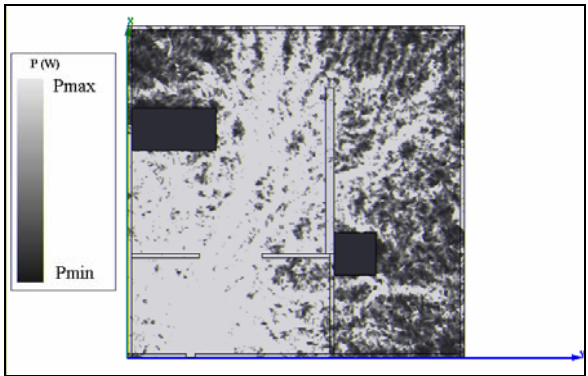

*fig 10 : Recouvrement de cellule et hand-over
- utilisation de la première source*

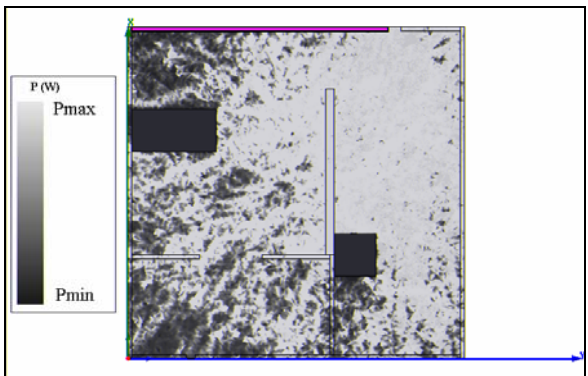

*fig 11 : Recouvrement de cellule et hand-over
- utilisation de la deuxième source*

3.6     Caractérisation des matériaux de construction

Cette partie doit permettre à l'étudiant de se familiariser avec les propriétés de réflexion et de transmission des principaux matériaux de construction afin qu'il puisse les « classer » en fonction de leur aptitude à atténuer ou à réfléchir les ondes électromagnétiques (Figure 12 et 13).

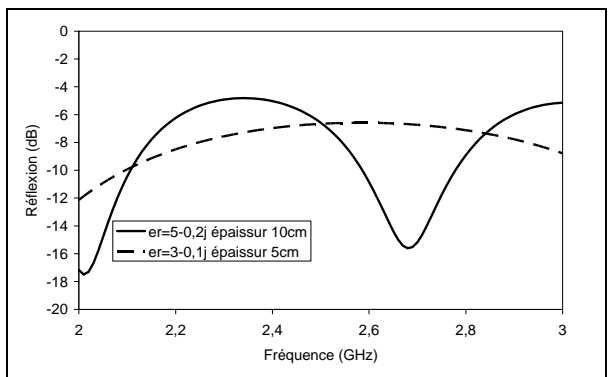

*fig 12 : Etude de matériaux en réflexion*

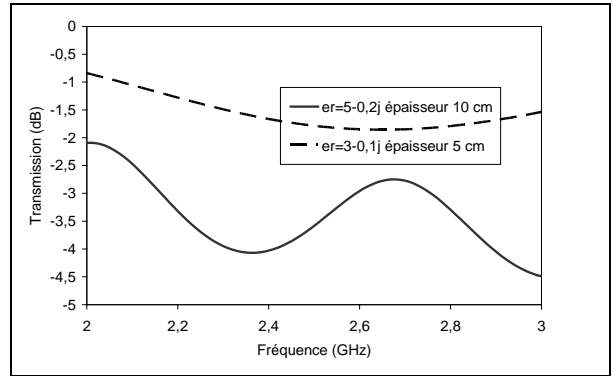

*fig 13 : Etude de matériaux en transmission*

Les résultats sont obtenus à l'aide d'un programme implanté sous le logiciel Matlab. Les résultats sont obtenus très rapidement. L'étudiant peut alors calculer les coefficients de réflexion et de transmission de différents matériaux   [4] (tableau 1) en faisant aussi varier leur épaisseur.

| Matériau | $\varepsilon'$ | $\varepsilon''/\varepsilon'. 10^4$ |
|---|---|---|
| Air | 1 | 0 |
| Béton,ciment | 2.4 | 7.8 |
| Bois | 2 | 300 |
| Plexiglass | 2.60 | 57 |
| Verre | 11 | 40 |
| polystyrène | 2.55 | 5 |
| Plastique | 5 | 500 |

*Tableau 1 : permittivité de quelques matériaux*

## 4    Conclusion

A travers des travaux pratiques expérimentaux (installation et configuration d'un point d'accès , étude de trames, mesure de champ rayonné) et des travaux pratiques utilisant des outils de simulation numérique des phénomènes électromagnétiques, l'étudiant peut acquérir l'ensemble des connaissances qui lui permettront d'installer un réseau WiFi efficace et sécurisé.

### Bibliographie

1. DEMONTOUX F., VIGNERAS-LEFEBVRE V Projet de travaux pratiques en télécommunication associant un banc de mesure à des simulations numériques
   Colloque CETSIS EEA, Clermont Ferrand, FRANCE, 10/01
2. Documentation logiciel HFSS-ANSOFT
3. A.Bossavit,C.Emson,I.D.Mayergoyz.   Méthodes numériques en électromagnétisme. Eyrolles. 1991 ; pp 149-245
4. A.R Von Hippel. Dielectric materials and applications. The technology press of M.I.T.. 1961